\documentclass[12pt,preprint]{aastex}

\begin{document}

\title{Non-Axisymmetric $g$-Mode and $p$-Mode Instability in\\ 
a Hydrodynamic Thin Accretion Disk}

\author{Li-Xin Li\altaffilmark{a,1},\, Jeremy
Goodman\altaffilmark{b},\, and Ramesh Narayan\altaffilmark{a}}
\affil{$^{a}$Harvard-Smithsonian Center for Astrophysics, Cambridge,
MA 02138, USA} \email{lli,rnarayan@cfa.harvard.edu}
\affil{$^{b}$Princeton University Observatory, Princeton, NJ 08544,
USA} \email{jeremy@astro.princeton.edu}

\altaffiltext{1}{Chandra Fellow}

\begin{abstract}
It has been suggested that quasi-periodic oscillations of accreting
X-ray sources may relate to the modes named in the title. We consider
non-axisymmetric linear perturbations to an isentropic, isothermal,
unmagnetized thin accretion disk.  The radial wave equation, in which
the number of vertical nodes ($n$) appears as a separation constant,
admits a wave-action current that is conserved except, in some cases,
at corotation.  Waves without vertical nodes amplify when reflected by
a barrier near corotation.  Their action is conserved.  As was
previously known, this amplification allows the $n=0$ modes to be
unstable under appropriate boundary conditions.  In contrast, we find
that waves with $n >0$ are strongly absorbed at corotation rather
than amplified; their action is not conserved.  Therefore,
non-axisymmetric $p$-modes and $g$-modes with $n>0$ are damped
and stable even in an inviscid disk.  This eliminates a
promising explanation for quasi-periodic oscillations in neutron-star
and black-hole X-ray binaries.

\end{abstract}


\section{Introduction}
\label{sec1}
In recent years, the study of quasi-periodic oscillations (QPOs) in
X-ray binaries has developed into a major field.  A variety of QPOs
have been observed in the variability power spectra of neutron-star
and black-hole X-ray binaries \citep[and references
therein]{kli00,rem02}, and the observations have revealed a rich
phenomenology.

Of particular interest are the ``kilohertz QPOs,'' which have
frequencies of several hundreds of Hz to occasionally more than $10^3$
Hz.  Because of their high frequencies, these QPOs must be produced by
processes close to the accreting mass.  However, since they have been
seen in both neutron-star and black-hole X-ray binaries, it appears
that the oscillations are not associated with the surface of the
accreting object.  Instead, it is generally believed that the
kilohertz QPOs originate in the accretion flow surrounding the central
mass.

A detailed understanding of the oscillation modes of accretion disks
could in principle allow observations of QPOs to be used to test
strong gravity in the vicinity of compact objects (e.g., Stella \&
Vietri 1998; Stella, Vietri \& Morsink 1999).  Observations could also
be ``inverted'' to measure relativistic parameters of the accreting
mass, such as the spin of a black hole (Nowak et al. 1997; Wagoner,
Silbergleit \& Ortega-Rodriguez 2001).  But such applications require
a robust method of associating different observed QPO frequencies with
specific disk modes and of calculating the frequencies of those modes
from first principles.  

In what follows, we focus on $g$-modes and $p$-modes in hydrodynamic
thin disks.  Following Silbergleit, Wagoner, \& Ortega-Rodr\'{\i}guez
(2001), Kato (2002), and references therein, we distinguish between
the two kinds of modes primarily in terms of where most of the wave
action is concentrated.  If the bulk of the action is near corotation,
we call it a $g$-mode, and if the action is mostly away from
corotation, we call it a $p$-mode.

In an important paper, Okazaki, Kato, \& Fukue (1987) showed that
axisymmetric $g$-modes are trapped in the inner regions of a
relativistic disk, where the epicyclic frequency $\kappa$ reaches a
maximum.  This idea was exploited by Nowak \& Wagoner (1991, 1992) and
a number of other workers \citep{per97,sil01,abr01} who worked out the
physical properties of these and related modes (see Kato, Fukue \&
Mineshige 1998, Wagoner 1999, Kato 2001a for reviews).  The modes are
not dynamically unstable, however. \citet{ort00} claim that viscosity
destabilizes $p$- and $g$- modes, but we limit ourselves to inviscid
disks.

Recently, Kato (2001b, 2002) claimed to demonstrate that {\it
non-axisymmetric} $g$-modes are trapped between two forbidden zones
that lie on either side of the corotation radius.  Kato further argued
that these modes are highly unstable.  This is very interesting
because the observed kilohertz QPOs often have quite a large amplitude
of flux variations, suggesting that the corresponding disk modes are
probably dynamically unstable.  This makes any disk modes that are
dynamically unstable to be of great interest for the QPO problem.
Although in a later paper, Kato (2003) has withdrawn the claim of an
instability, his work nevertheless suggests that it might be fruitful
to explore dynamical instabilities in non-axisymmetric disk modes.
This is the motivation behind the present paper.

The work presented here is particularly influenced by the papers of
Goldreich \& Narayan (1985, henceforth GN) and Narayan, Goldreich, \& 
Goodman (1987, henceforth NGG).
These authors studied perturbations in a particular simplified fluid
system called the shearing sheet.  They showed that non-axisymmetric
modes with no vertical nodes ($n=0$ in the notation used in this
paper) can be dynamically unstable.  They further
demonstrated that the driving force for the instability is a wave
amplifier inside the system.  These papers were in turn influenced
by work on spiral wave amplification in stellar disks,
especially \citet{Mark76}.

We use the physical understanding obtained from the above work to
guide our present investigation.  In \S2 we consider a thin accretion
disk (which is more general than the shearing sheet), and derive a
wave equation for linear non-axisymmetric perturbations and an
associated conserved current.  In \S3 we explore the nature of
the singularities in the wave equation associated with the corotation
resonance and the Lindblad resonances.  In \S4 we consider the WKB
limit of the wave equation and identify the basic properties of
ingoing and outgoing waves in various regions of the disk.  In \S5 we
consider the interaction of waves with various barriers associated
with the Lindblad and/or corotation resonance.  We show that waves
with $n=0$ are amplified when they reflect off the corotation barrier.
This confirms the result obtained by GN and NGG.
However, when we repeat the analysis for waves
with $n>0$ (waves with one or more nodes in the vertical direction),
we find that there is no amplifier anywhere in the system, either at
the Lindblad resonances (\S5) or at corotation (\S6).  Indeed, we show
that corotation acts as a severe absorber of waves.  Based on these
results, we conclude in \S7 that there are no dynamically unstable
$p$- or $g$-modes with $n>0$ in a thin disk.  Appendix A presents some
numerical results which help to extend the analysis to the non-WKB
regime.

\section{Wave Equation for Non-Axisymmetric Perturbations to a Thin Disk}
\label{sec2}

We consider an axisymmetric thin disk.  Because of the symmetry of the
unperturbed system, we assume without loss of generality that the
linear perturbations are proportional to $\exp[i (-\omega t + m\phi)]$
in cylindrical coordinates $(r,\phi,z)$.  Here, $m$ is an integer and
$\omega$ is the mode frequency, which may be either real or complex.
We write the linear perturbations to the mass density and velocity as
\begin{eqnarray}
	\rho(t,r,\phi,z) &=& \rho_0(r,z) + \rho_1(r,z) \exp[i (-\omega t + 
		m\phi)] \;, \\
	v_r(t,r,\phi,z) &=& u(r,z) \exp[i (-\omega t + m\phi)] \;, \\
	v_\phi(t,r,\phi,z) &=& \Omega(r) r + v(r,z) \exp[i (-\omega t + 
		m\phi)] \;, \\
	v_z(t,r,\phi,z) &=& w(r,z) \exp[i (-\omega t + m\phi)] \;,
\end{eqnarray}
where $\rho_0$ is the unperturbed mass density, $\Omega$ is the
unperturbed angular velocity, and $\rho_1$, $u$, $v$, and $w$ are
first-order perturbations to the mass density and the velocity field.
Note that $\partial\Omega/\partial z=0$ since we take the disk to be 
isentropic. Following Kato, we use Newtonian equations.  Although general
relativity may introduce additional instabilities, many of its effects
can be mimicked by appropriate radial variations of the structural
disk frequencies $\Omega$, $\kappa$, and $\Omega_\perp$, the latter
two of which are defined below.

For simplicity, we assume that the disk has an isothermal equation of
state,
\begin{eqnarray}
	p = \rho c_{\rm s}^2 \;,
\end{eqnarray}
where $p$ is the gas pressure and $c_{\rm s}$ is the sound speed which
we take to be a constant.  The equilibrium structure of a sufficiently
thin isothermal disk then takes the form
\begin{eqnarray}
	\rho_0(r,z) = \rho_{00}(r) \exp\left(-\frac{z^2}{2h^2}\right) \;, 
\end{eqnarray}
where $h = h(r) = c_{\rm s}/\Omega_\perp(r)$ is the half-thickness of
the disk, and $\Omega_\perp$ is the vertical disk frequency, which may
be different from the orbital angular velocity due to strong radial
pressure gradients in a Newtonian disk (unlikely for a thin disk), or
to relativistic gravity.

The isothermal equation of state allows an almost exact separation of 
variables, as shown below. This simplification is well worth the
loss of generality, especially since the main focus of our paper is on 
the corotation resonance. In the vicinity of the corotation resonance, 
in fact, the frequency $\sigma$ (defined below) is
small compared to the reciprocal of the sound crossing time across
the thickness of the disk; consequently, motions near corotation
are noncompressive, so that the choice among different isentropic equations
of state is not important.  Far from corotation
($|r-r_{\rm c}|\gtrsim r/m$), however, 
non-isothermal equilibria concentrate waves toward
the disk surface where they may be more easily dissipated
\citep[and references therein]{BOLP02}.

In terms of the enthalpy perturbation
\begin{eqnarray}
	Q = \frac{\delta p}{\rho_0} \;,
\end{eqnarray}
the first-order perturbation to the continuity equation is
\begin{eqnarray}
	\frac{1}{r\rho_0}\frac{\partial}{\partial r}(r \rho_0 u) + 
		\frac{im}{r} v+ \frac{1}{\rho_0}\frac{\partial}
		{\partial z}(\rho_0 w) = \frac{i\sigma}
		{c_{\rm s}^2} Q \;,
	\label{cont}
\end{eqnarray}
where 
\begin{eqnarray}
	\sigma \equiv \omega - m \Omega 
\end{eqnarray}
is the frequency of the perturbation in the local corotating frame of
the disk. The first-order perturbations to the radial, azimuthal, and
vertical momentum equations lead to
\begin{eqnarray}
	i \sigma u + 2\Omega v &=& \frac{\partial Q}{\partial r} \;, 
		\label{rmom}\\
	\frac{\kappa^2}{2\Omega}\, u - i\sigma v &=& - \frac{im}{r} Q \;,
		\label{amom}\\
	i\sigma w &=& \frac{\partial Q}{\partial z} \;,
		\label{zmom}
\end{eqnarray}
respectively, where
\begin{eqnarray}
	\kappa \equiv \sqrt{2\Omega \left(2\Omega + r \frac{d\Omega
		}{d r}\right)} 
\end{eqnarray}
is the epicyclic frequency of the flow in the $r\phi$-plane. 

Solving for $u$ and $v$ from equations~(\ref{rmom}) and (\ref{amom}),
and $w$ from equation~(\ref{zmom}), and substituting the results into
equation~(\ref{cont}), we obtain a second-order partial differential
equation for $Q$
\begin{eqnarray}
	\frac{\sigma^2}{\rho_0 r}
	\frac{\partial}{\partial r}\left(\frac{\rho_0 r}{D} \frac{\partial
		Q}{\partial r}\right) - \left[\frac{\sigma^2}
		{c_{\rm s}^2} +\left(\frac{m\sigma}{r}\right)^2 \frac{1}
		{D} + \frac{2 m\sigma}{\rho_0 r}\frac{\partial}{\partial 
		r}\left(\frac{\rho_0\Omega}{D}\right)\right] Q 
		= \frac{1}{\rho_0}\frac{\partial}{\partial z}\left(\rho_0 
		\frac{\partial Q}{\partial z}\right) \;,
	\label{pdeq}
\end{eqnarray}
where
\begin{equation}
D \equiv \kappa^2 -\sigma^2 .
\label{D}
\end{equation}

Equation~(\ref{pdeq}) is most easily analysed by separation of
variables in $r,z$, as discussed by Kato (2001a).  Since $dh/dr \sim
h/r$ is small for a thin disk, we neglect the dependence of $h$ on $r$
and introduce a coordinate $\eta \equiv z/\left(\sqrt{2}h\right)$ to 
replace $z$. Then,
writing $Q(r,\eta) = Q_r(r) Q_\eta(\eta)$ and substituting in
equation~(\ref{pdeq}), we can show that $Q_r(r)$ and $Q_\eta(\eta)$
satisfy
\begin{eqnarray}
	\frac{d}{d r}\left(\frac{\rho_{00}r}{D} \frac{dQ_r}{dr}
		\right) + \frac{\rho_{00} r}{\sigma^2}\left[
		\frac{n\Omega_\perp^2 -\sigma^2}{c_{\rm s}^2} -
		\frac{m^2 \sigma^2}{r^2 D} - \frac{2 m\sigma}{\rho_{00} 
		r}\frac{d}{dr}\left(\frac{\rho_{00}\Omega}{
		D}\right)\right] Q_r 
		= 0 \;,
	\label{deq1}
\end{eqnarray}
\begin{eqnarray}
	\frac{d^2 Q_\eta}{d\eta^2}- 2 \eta \frac{dQ_\eta}{d\eta}
		+ 2 n Q_\eta = 0 \;,
	\label{deqz2}
\end{eqnarray}
where $n$ is a constant.

Equation~(\ref{deqz2}) has the form of the Hermite equation. In order
for the energy density of perturbations to be bound as
$\eta\rightarrow\pm\infty$, $n$ must be a non-negative integer
\citep{oka87}: $n = 0, 1, 2, ...$. Then the solutions to 
equation~(\ref{deqz2}) are given by the Hermite polynomials: $Q_\eta
= H_n(\eta)$. The integer $n$ determines the
number of nodes in the vertical direction. An even (odd) $n$
corresponds to an even (odd) mode of oscillation.  The mode with $n=0$
has $Q_\eta =$ constant, and has no motions in the vertical direction
according to equation (\ref{zmom}).

Equation~(\ref{deq1}), with $n = 0, 1, 2, ...$, is the wave equation
for linear perturbations of an isothermal thin 
disk.  \citet{kat01a} obtained a similar equation for more general
equations of state by neglecting slowly radially varying terms.

The quantity
\begin{eqnarray}
	J \equiv \frac{i}{2 W} \left(Q_r^*\, \frac{dQ_r}{dr} -
		Q_r\, \frac{dQ_r^*}{dr}\right) \;, \hspace{1cm}
	W \equiv \frac{D}{\rho_{00} r} \;,
	\label{current}
\end{eqnarray}
where the asterisks denote complex conjugates, represents a
current of wave action. It can be checked that this
current is conserved, that is,
$dJ/dr = 0\,$, wherever equation~(\ref{deq1}) is nonsingular
and $\omega$ is real.  The current plays a prominent role in all
of the analysis presented below.  
The conservation law for wave action can be extended to complex
frequencies $\omega=\omega_R+i\omega_I$,
\[
-\frac{dJ}{dr}= 2\omega_I\rho_a \;,
\]
where $\rho_a$ is a real quantity proportional to $|Q_r|^2$ representing
the density of wave action (NGG).
The role of the singularities is more important in the present context.
We shall show that $J$ is conserved at all singularities except
corotation, and even there it is conserved when $n=0$.

\section{Corotation and Lindblad Singularities in the Wave Equation}
\label{sec3}

When $\sigma$ is real (i.e., $\omega$ is real), the wave
equation~(\ref{deq1}) contains two types of singularities: one is the
corotation singularity given by the condition $\sigma = 0$, the other
is the Lindblad singularities given by the condition $D = 0$.  The
former occurs at the corotation radius $r_{\rm c}$ and the latter at
the Lindblad radii $r_{\rm L}$.

\subsection{Corotation singularity: $n = 0$}
\label{sec3.1}

We begin by first discussing the case $n=0$, which corresponds to the
problem analysed by GN and NGG.  They considered the special case of the 
shearing sheet,
whereas we consider here a more general disk; however, near
corotation, the two problems are very similar.  For $r$ close to
$r_{\rm c}$, the wave equation becomes
\begin{eqnarray}
	\frac{d^2 Q_r}{dr^2} -\frac{A_{\rm c}Q_r}{r_{\rm c}(r-r_{\rm c})} 
		= 0 \;, \hspace{1cm}
		A_{\rm c}\equiv \left[\frac{2}{-d\ln\Omega/d\ln r}
		\frac{d}{d\ln r}\ln\left(\frac{\rho_{00}\Omega}{D}\right)
		\right]_{r=r_{\rm c}} \;,
	\label{eqrc2}
\end{eqnarray}
where we have used $\sigma = m \left(-d\Omega/dr\right)_{r=r_{\rm c}}
(r-r_{\rm c})$ near $r = r_{\rm c}\,$. 
Assuming that $Q_r \propto (r-r_{\rm c})^{\beta_1}$ near $r = r_{\rm
c}$, the leading term in equation~(\ref{eqrc2}) gives rise to
\begin{eqnarray}
	\beta_1 (\beta_1 -1) (r-r_{\rm c})^{\beta_1 -2} = 0 \;,
\end{eqnarray}
which requires $\beta_1 = 0$, or $1$. Thus, near the corotation radius
the solutions to the wave equation are
\begin{eqnarray}
	Q_r \propto 1\;, \hspace{0.8cm} r-r_{\rm c} \;.
	\label{solrc2}
\end{eqnarray}

Equation~(\ref{solrc2}) implies that the solutions are analytic near
$r = r_{\rm c}$.  Correspondingly, as can be easily checked, the
current defined by equation~(\ref{current}) is conserved
across $r_{\rm c}$, i.e., 
\begin{eqnarray}
	\frac{J_{{\rm c}-}}{J_{{\rm c}+}} = 1 \;, \qquad
	J_{{\rm c}\pm} \equiv J\left(r-r_{\rm c}=0^\pm\right) \;. 
\end{eqnarray}
Therefore, when $n = 0$, there is no real singularity at corotation, 
only an apparent singularity.  Indeed, for the model of a shearing 
sheet, NGG
have shown that the corotation singularity does not appear at all if
the perturbation equation is written in terms of the azimuthal
velocity rather than $Q_r$.

\subsection{Corotation singularity: $n > 0$}
\label{sec3.2}

When $n>0$, the wave equation near the corotation radius becomes
\begin{eqnarray}
	\frac{d^2 Q_r}{dr^2} + \frac{n b^2 Q_r}{(r-r_{\rm c})^2} 
		= 0 \;, \hspace{1cm}
		b\equiv \left.\frac{\Omega_\perp \kappa}{mc_{\rm s} 
			(-d\Omega/dr)}\right|_{r = r_{\rm c}} \;.
	\label{eqrc}
\end{eqnarray}
The solutions to equation~(\ref{eqrc}) are
\begin{eqnarray}
	Q_r \propto (r-r_{\rm c})^{\frac{1}{2}\pm 2iq} \;, \hspace{1cm}
		q\equiv \frac{1}{2} \sqrt{n b^2 -\frac{1}{4}} \;.
	\label{solrc}
\end{eqnarray}
Due to the presence of the factor $(r-r_{\rm c})^{1/2}$, the solutions
are not analytic at $r = r_{\rm c}$.  Thus, we have a real singularity.  
Note that, for a thin disk, we usually have $b\sim r/mh\gg 1$ unless 
$m$ is $\ga r/h\,$, so generally we expect $q\gg 1$.

Substituting the two solutions in (\ref{solrc}) into
equation~(\ref{current}), we can calculate the corresponding current
density $J_{{\rm c}+}$ defined in the previous subsection. Making the
analytic continuation $r-r_{\rm c} \rightarrow (r-r_{\rm c})e^{\pi i}$
to the solutions in equation~(\ref{solrc}) and substituting the
results into equation~(\ref{current}), we can calculate the
corresponding current $J_{{\rm c}-}$. We find that the current is not 
conserved across corotation:
\begin{eqnarray}
	\left|\frac{J_{{\rm c}-}}{J_{{\rm c}+}}\right| = e^{\mp 4\pi q} \;,
\end{eqnarray}
where the upper/lower sign corresponds to the upper/lower sign in 
equation~(\ref{solrc}), respectively.

This analysis shows that the behavior near corotation is very
different for $n=0$ and $n>0$.  This is one of two major differences
between the two cases, the other being the geometry of permitted and
forbidden zones for waves, discussed in \S4 and displayed in Figures~1
and 2.

\subsection{Lindblad singularity}
\label{sec3.3}

The Lindblad singularity occurs at the radii $r_{\rm L}$ where $D=0$.
Near this singularity, we have $\sigma \approx \pm \kappa$ and
$D\approx D_{\rm L}^\prime (r -r_{\rm L})$, assuming that $D_{\rm
L}^\prime \equiv (dD/dr)_{r =r_{\rm L}} \neq 0\,$. Then, assuming that
$Q_r \propto (r-r_{\rm L})^{\beta_2}$ near $r = r_{\rm L}$, the leading
term in equation~(\ref{deq1}) gives rise to
\begin{eqnarray}
	\beta_2 (\beta_2 -2) (r-r_{\rm L})^{\beta_2 -3} = 0 \;,
\end{eqnarray}
which requires $\beta_2 = 0$, or $2$. Thus, when $D_{\rm L}^\prime\neq
0\,$, the two solutions to the wave equation are locally
\begin{eqnarray}
	Q_r \propto 1\;, \hspace{0.8cm} (r-r_{\rm L})^2 \;.
\end{eqnarray}
Both solutions are clearly analytic.

More generally, if $D$ is an analytic function of $r$ near the
Lindblad radius (as it is in our problem), the solutions to the wave
equation must also be analytic there. It can be checked that the
current defined by equation~(\ref{current}) is then conserved
across the Lindblad singularity.  Therefore, the Lindblad resonance is
not a true singularity but only an apparent singularity
\citep[NGG;][]{kat02}. This is self-evident when the second-order
equation (\ref{eqrc2}) is replaced by the equivalent system of
first-order equations (\ref{cemat}).  Indeed, if one
considers wave equations for other perturbed quantities (e.g., radial
velocity), the Lindblad singularity gives way to other singular terms,
which occur at different radii and are again not real singularities
but only apparent. The corotation singularity, on the other hand,
persists for any fluid variable one selects and is a genuine
singularity when $n >0$.

\vspace{0.5cm}

In summary, when $n>0$ the corotation resonance is an intrinsic
singularity in the wave equation, where the conservation of the
current breaks down.  When $n = 0$, the corotation resonance
is only an apparent singularity; the solutions are well-behaved there
and the current is conserved across the resonance. The
Lindblad resonance is never a real singularity; the solutions are
well-behaved and the current is conserved across the Lindblad
resonance for all values of $n$.

\section{WKB Solutions to the Wave Equation in Permitted Regions}
\label{sec4}

In the WKB regime, where the wavelength is much smaller than the length scale
over which the potential in the wave equation varies, the wave 
equation~(\ref{deq1}) can be approximated by
\begin{eqnarray}
	\frac{d^2 Q_r}{d r^2}\, -\frac{d\ln W}{dr}\frac{dQ_r}{dr}
             +k_r^2(r)\, Q_r = 0 \;,
	\label{wkbeq} 
\end{eqnarray}
where
\begin{eqnarray}
	k_r^2(r) \equiv \frac{\left(\sigma^2- n \Omega_\perp^2\right)
		(\sigma^2 -\kappa^2)}{c_{\rm s}^2\, \sigma^2} \;,
	\label{k2}
\end{eqnarray}
In the last step, we have assumed that $m h/r \ll 1$.
Equation~(\ref{wkbeq}) is Kato's (2001b, 2002) wave equation, except
that we have retained the subdominant term involving the Wronskian $W$
(eq.~[\ref{current}]). This term is crucial because it influences the
sign of the wave action and hence the amplification of waves.

When the condition
\begin{eqnarray}
	\left|\frac{1}{k_r^2}\frac{dk_r}{dr}\right| 
	\sim \left|\frac{1}{k_r r}\right| \ll 1
	\label{wkbcon}
\end{eqnarray}
is satisfied, the wave equation~(\ref{wkbeq}) can be solved with the WKB 
approximation (see, e.g., Merzbacher 1998):
\begin{eqnarray}
	Q_{r,{\rm WKB}} \approx \sqrt{\frac{W}{k_r}}\, \exp\left(\pm i
		\int^r k_r dr\right) \;.
	\label{qrwkb}
\end{eqnarray}
The solution with the ``$+$'' sign in the exponential corresponds to
wave-vector $+k_r$, and the solution with the ``$-$'' sign corresponds to 
wave-vector $-k_r$.

\begin{figure}
\epsscale{1}
\plotone{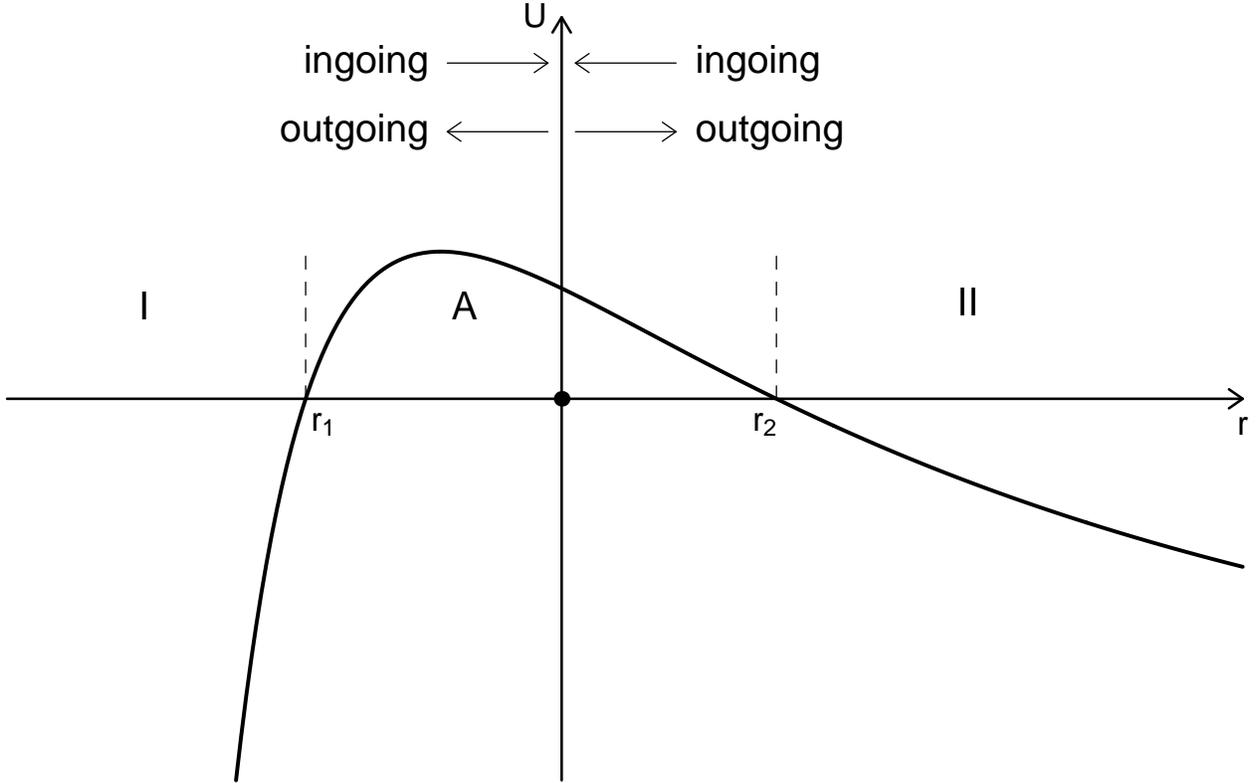}
\caption{The potential $U(r)\equiv -k_r^2(r)$ in the wave 
equation~(\ref{wkbeq}) for the case when $n = 0$. There are two permitted 
regions and one forbidden region along the $r$-axis. The two permitted 
regions are: I ($r<r_1$) and II ($r>r_2$), where $r_{1,2}$ are defined by 
$\sigma=\mp\kappa$. The boundaries of the regions are marked by the two 
vertical dashed lines. The forbidden region is: A ($r_1<r<r_2$), which
contains the corotation radius as marked by the dot. (When $n=0$
the corotation radius is not a singularity in the wave 
equation~[\ref{wkbeq}].) The directions of ingoing and outgoing waves are 
shown with horizontal arrows, which are defined relative to the corotation 
radius. The relations among the directions of the current, the 
wave-vector, and the group velocity in each permitted region are 
summarized in Table~\ref{tab1}. 
\label{fig1}}
\end{figure}

Within the WKB approximation, permitted regions for waves are defined
by the requirement $k_r^2(r) > 0$, and forbidden regions by $k_r^2(r)
< 0$.  Consider first the simpler case of $n=0$.  Here, the region of
the disk in which $\sigma^2<\kappa^2$ is forbidden.  This region
extends from the inner Lindblad radius ($\sigma=-\kappa$) to the outer
Lindblad radius ($\sigma=\kappa$), straddling the corotation radius
($\sigma=0$).  It is shown as forbidden region A in Figure~\ref{fig1}.  
On either side of A, there are two permitted regions: region I with
$\sigma<-\kappa$, and region II with $\sigma >\kappa$.

\begin{figure}
\epsscale{1}
\plotone{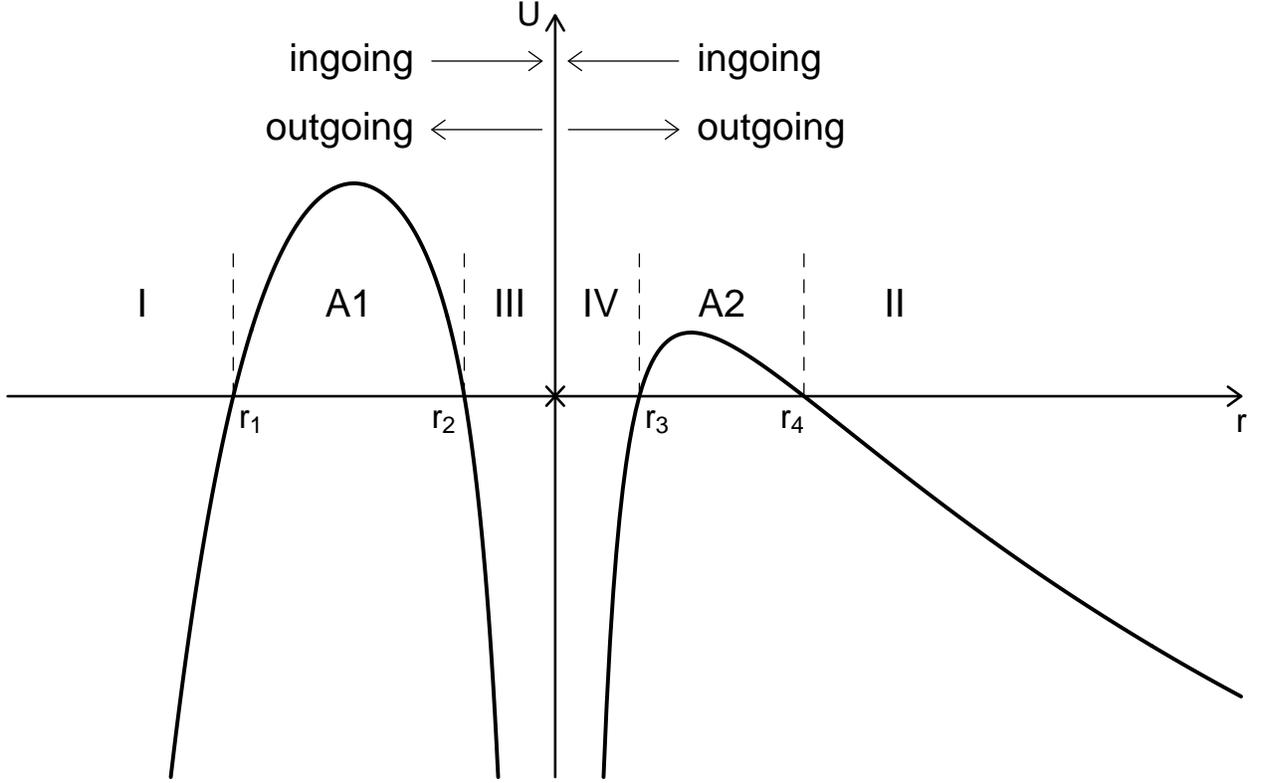}
\caption{The potential $U(r)\equiv -k_r^2(r)$ in the wave 
equation~(\ref{wkbeq}) for the case when $n > 0$.  There is an intrinsic 
singularity at the corotation radius at $r = r_{\rm c}$ where $\sigma=0$ and 
the potential is infinitely deep, as indicated by the cross-sign. There 
are four permitted regions and two forbidden regions along the $r$-axis.  
The permitted regions are: I ($r<r_1$), II ($r>r_4$), III ($r_2<r<r_{\rm c}$), 
and IV ($r_{\rm c}<r<r_3$), where $r_{1,4}$ are defined by $\sigma=\mp
\max\left(\sqrt{n} \Omega_\perp,\kappa\right)$, and $r_{2,3}$ are defined by 
$\sigma=\mp\min\left(\sqrt{n}\Omega_\perp,\kappa\right)$. The boundaries 
of the regions are marked by the four vertical dashed lines plus the 
$U$-axis. The forbidden regions are: A1 ($r_1<r<r_2$), and A2 ($r_3<r<r_4$).
The directions of ingoing and outgoing waves are shown with horizontal 
arrows, which are defined relative to the corotation radius. The 
relations among the directions of the current, the wave-vector, 
and the group velocity in each permitted region are summarized in 
Table~\ref{tab1}. 
\label{fig2}}
\end{figure}

When $n > 0$, things are more complicated.  Now, there are four
permitted regions (see Fig.~\ref{fig2}):
\begin{itemize}
\item[--]{Region I: the region with $\sigma < -\max\left(\sqrt{n} 
\Omega_\perp, \kappa\right)\,$;} 
\item[--]{Region II: the region with $\sigma > \max\left(\sqrt{n} 
\Omega_\perp, \kappa\right)\,$;}
\item[--]{Region III: the region with $-\min\left(\sqrt{n} \Omega_\perp,
\kappa\right) < \sigma < 0\,$;}
\item[--]{Region IV: the region with $0 < \sigma <\min\left(\sqrt{n} 
\Omega_\perp, \kappa\right)\,$.}
\end{itemize}
Regions I and III have $\sigma < 0$ and are on the left-hand side
of corotation.  These two zones are separated from each other by a
forbidden zone (barrier A1, see Fig.~\ref{fig2}).  Regions II and
IV have $\sigma > 0$ and are on the right-hand side of corotation.
They are again separated from each other by a forbidden zone (barrier
A2).  Regions III and IV are separated from each other by the
corotation resonance, which is not a true barrier.  In contrast to the
$n=0$ case, here the region around corotation is permitted.

For a WKB solution with a wave-vector $k_r$ (i.e., the solution in
eq.~[\ref{qrwkb}] with the ``$+$'' sign in the exponential), the
corresponding current is
\begin{equation}\label{Jwkb}
        J_{\rm WKB} \approx \frac{k_r}{-W} \left| Q_{r,{\rm WKB}}
                \right|^2 \;,
	\label{jwkb}
\end{equation}
where we have used the fact that in the permitted regions $k_r$ is
real so that $k_r^* = k_r$. From the definition of $W$
(eq.~[\ref{current}]), the sign of $W$ is determined by the sign of $D
= \kappa^2 -\sigma^2$. Hence, the current and the wave-vector
have the same signs in regions where $\sigma^2 > \kappa^2$ (regions I
and II, see Table 1), but opposite signs where $\sigma^2 < \kappa^2$
(regions III and IV).

We now define an ``outgoing'' wave as one that moves away from
corotation, i.e., towards larger $|\sigma|$, and an ``ingoing'' wave
as one that moves towards corotation (see Figs.~\ref{fig1} and
\ref{fig2}).  The direction of motion is determined by the sign of the
group velocity $v_{{\rm g}r}$ rather than the wave-vector $k_r$.

By definition, $v_{{\rm g}r} \equiv (\partial k_r/
\partial\omega)^{-1} = (\partial k_r/\partial\sigma)^{-1}$, so that
\begin{equation}
        k_r v_{{\rm g}r} = \sigma \left(\frac{\partial\ln k_r^2}
		{\partial\ln \sigma^2}\right)^{-1} \;.
        \label{vg}
\end{equation}
From equation~(\ref{k2}) we have
\begin{equation}
        \frac{\partial\ln k_r^2}{\partial\ln \sigma^2} = 
		\frac{\sigma^4 - n \Omega_\perp^2 \kappa^2}
                {\left(\sigma^2 -n \Omega_\perp^2\right)
		\left(\sigma^2 - \kappa^2\right)} \;,
\end{equation}
and therefore
\begin{equation}
        k_r v_{{\rm g}r} = 
		\frac{c_{\rm s}^2\,k_r^2\sigma^3}{\sigma^4 - n 
		\Omega_\perp^2 \kappa^2} \;.
        \label{dkr}
\end{equation}
It is then easily verified that $k_r v_{{\rm g}r} >0$ in regions II
and III, and $k_r v_{{\rm g}r} <0$ in regions I and IV. This allows us
to identify whether a particular wave in a given permitted region is
ingoing or outgoing.

Knowing the signs of $k_r J$ and $k_r v_{{\rm g}r}$ in each permitted
region, we are able to determine the sign of $v_{{\rm g}r} J$ in that
region (Table~\ref{tab1}).  This is an important quantity, as we show
in the next section.

Finally, we briefly comment on the locations of the four permitted
regions in a real disk. From their definitions, if all four regions I,
II, III, and IV exist, they should appear in the following order with
increasing disk radius (as in Fig.~\ref{fig2}): region I, region III,
region IV, and region II, since $\sigma$ is an increasing function of
radius. Thus we expect region I to be near the inner boundary of the
disk ($r \approx r_{\rm in}$), region II to be at large radii ($r \gg
r_{\rm c}$), and regions III and IV to lie in between, on either side
of corotation.

\section{Wave Amplification/De-amplification at the Barriers}
\label{sec5}

As explained in \S1, a dynamical instability is possible if there is a
wave amplifier in the system.  In the case of $n=0$, such an amplifier
is present at the corotation barrier (GN,
NGG), as we now demonstrate using the results derived
in \S{\ref{sec4}}.  Figure~\ref{fig1} shows that, when $n = 0$, there
are only two permitted regions in the disk: region I near the inner
boundary of the disk, and region II at large radii.  These two regions
are separated from each other by barrier A, which includes the
corotation and Lindblad resonances.  However, as shown in
\S\S\ref{sec3.1} and \ref{sec3.3}, neither the corotation
resonance nor the Lindblad resonances are real singularities, 
and the current defined by equation~(\ref{current}) is
conserved throughout the disk.

Consider now a wave incident on the potential barrier from region I. 
The interaction of this
ingoing wave with the potential barrier will produce a reflected
outgoing wave in region I and a transmitted outgoing wave in region
II. Because of the conservation of the current, we have the
following relation among the currents of the three waves:
\begin{eqnarray}
	J_{\rm in}^{\rm (I)} + J_{\rm out}^{\rm (I)} =  
		J_{\rm out}^{\rm (II)} \;.
	\label{current12}
\end{eqnarray}

From Table~\ref{tab1}, in region II the group velocity and the current
density of the wave have the same sign. Since an outgoing wave in
Region II has a positive group velocity (the wave goes away from
corotation towards larger $r$), the corresponding current
$J_{\rm out}^{\rm (II)}$ must be positive.  Thus, the right-hand side
of equation (\ref{current12}) is positive.  In region I, the group
velocity and the current of the wave have opposite signs.  The
ingoing wave here has a positive group velocity (it is moving from
small values of $r$ towards $r_{\rm c}$), while the outgoing wave has
a negative group velocity.  Thus, $J_{\rm in}^{\rm (I)} <0$ and
$J_{\rm out}^{\rm (I)} >0$. Then, equation~(\ref{current12}) is
equivalent to
\begin{eqnarray}
	\left|J_{\rm out}^{\rm (I)}\right| = \left|J_{\rm in}^{\rm (I)}
		\right| + \left|J_{\rm out}^{\rm (II)}\right| \;.
	\label{current12a}
\end{eqnarray}

We define the \emph{gain} $G$ to be the absolute value of the reflection 
coefficient. Equation~(\ref{current12a}) implies that the gain
\begin{equation}
G \equiv \left|{J_{\rm reflected} \over J_{\rm incident}}\right| =\left|
{J_{\rm out}^{\rm (I)} \over J_{\rm in}^{\rm (I)}}\right|
>1, \qquad n=0. \label{G1}
\end{equation}
In other words, for $n=0$, an incident wave is reflected by the
potential barrier with a larger amplitude, and the potential barrier
behaves as an amplifier.  If, in addition, the inner edge of the disk
behaves like a near-perfect reflector, then we have the makings of an
instability.  (This would be a $p$-mode disk instability since the
wave action is concentrated away from corotation.)

The corotation amplifier works equally well if a wave
is incident on the potential barrier from region II, as is
easily verified.  In this case, to have an instability, the outgoing
wave in region II must be somehow reflected.  The reflection is
unlikely to be from the distant outer edge of the disk.
Perhaps an inhomogeneity in the disk might provide the
necessary reflection, but the topic is beyond the scope of this paper.
For the purposes of this paper, the key point is that, when $n=0$, the
system has a wave amplifier and, therefore, can potentially have
unstable modes.

The situation changes dramatically when $n>0$.  Now, there are four
permitted regions in the disk, I, II, III, IV, and two barriers,
A1, A2.  Let us consider the region to the right of corotation
(i.e., $\sigma >0$). Assume that an outgoing wave is incident on the
barrier A2 from region IV. The interaction of this outgoing wave
with the barrier produces a reflected ingoing wave in region IV and a
transmitted outgoing wave in region II.  We have shown earlier that
there is no real singularity at the barrier, and so the current
density must be conserved from region IV to region II.  We then have
\begin{eqnarray}
	J_{\rm out}^{\rm (IV)} + J_{\rm in}^{\rm (IV)} =  
		J_{\rm out}^{\rm (II)} \;.
	\label{current42}
\end{eqnarray}

From Table~\ref{tab1} we see that in region IV the group velocity and
the current have the same sign. Since in region IV an outgoing
wave has a positive group velocity and an ingoing wave has a negative
group velocity, we have $J_{\rm out}^{\rm (IV)} > 0$ and $J_{\rm
in}^{\rm (IV)} < 0$. Similarly, $J_{\rm out}^{\rm (II)} > 0$.  Thus,
equation~(\ref{current42}) can be rewritten as
\begin{eqnarray}
	\left|J_{\rm in}^{\rm (IV)}\right| = \left|J_{\rm out}^{\rm (IV)}
		\right| - \left|J_{\rm out}^{\rm (II)}\right| \;.
	\label{current42a}
\end{eqnarray}
Equation~(\ref{current42a}) implies that the gain in this
case is always less than unity:
\begin{equation}
G = \left|{J_{\rm reflected} \over J_{\rm incident}}\right| 
=\left|{J_{\rm in}^{\rm (IV)} \over J_{\rm out}^{\rm (IV)}}\right|
<1, \qquad n>0. \label{G2}
\end{equation}
In other words, a wave that is incident on barrier A2 from region
IV is always reflected with a smaller amplitude.  It can be easily
checked that this statement is true also for a wave incident on
barrier A2 from region II, and also for waves incident on barrier A1
from either region I or region III.

We thus conclude that, for $n>0$, neither of the two barrier A1 and
A2 behaves like an amplifier; both barriers deamplify waves.  This
eliminates a promising mechanism for producing a dynamical
non-axisymmetric instability.  The only thing left to be checked is
whether corotation itself can amplify waves.  This is the topic of the
next section.

\section{Absorption at the Corotation Resonance}
\label{sec6}

To understand the role played by the corotation singularity when
$n>0$, we need to study the behavior of the solutions of the wave
equation~(\ref{deq1}) near $\sigma = 0$ (i.e., $r = r_{\rm c}$). In
particular, we would like to know for a wave incident on the
resonance, how much is transmitted to the other side and how much is
reflected.

As $|\sigma| \rightarrow 0$ (i.e., $r\rightarrow r_{\rm c}$), and
taking $n>0$, the wave equation becomes equation~(\ref{eqrc}), whose
two linearly independent solutions are given by equation~(\ref{solrc})
for $q\neq 0$. The wave-vectors corresponding to the two solutions are
\begin{eqnarray}
	k_r = \pm \frac{\sqrt{n}\,b}{r -r_{\rm c}} \;.
\end{eqnarray}
It can be checked that when $\sqrt{n}\, b\gg 1$ (i.e., $q\gg 1$) the 
solutions in equation~(\ref{solrc}) become the WKB solutions given by 
equation~(\ref{qrwkb}).

According to Table~\ref{tab1}, in region IV ($r> r_{\rm c}$) the
wave-vector and the group velocity have opposite signs. Thus, in
region IV the ingoing wave, which has a negative group velocity i.e. a
positive wave-vector, is given by $(r-r_{\rm
c})^{\frac{1}{2}+2iq}$. Now let us analytically continue this solution
into region III.  As discussed in NGG, the
continuation must be done such that the integration path in the
complex plane goes above the singularity at $\sigma = 0$ if the
solution is to correspond to an initial-value problem. That is, we
must take $r-r_{\rm c} \rightarrow \left| r-r_{\rm c}\right| e^{\pi
i}$. We then find that the solution transforms as follows:
\begin{equation}
	{\rm Region~IV \to Region~III}: \qquad (r-r_{\rm c})^{\frac{1}
		{2}+2iq} \rightarrow i e^{-2\pi q} \left|r- r_{\rm c}
		\right|^{\frac{1}{2}+2iq} \;. \label{io1}
\end{equation}

In region III ($r< r_{\rm c}$) the wave-vector and the group velocity
have the same sign. So, in region III the outgoing wave, which has a
negative group velocity i.e. a negative wave-vector, is given by
$(r-r_{\rm c})^{\frac{1}{2} +2iq} \propto \left|r-r_{\rm
c}\right|^{\frac{1}{2}+2iq}$. Thus, from equation~(\ref{io1}), the
ingoing wave in region IV becomes a purely outgoing wave with a
reduced amplitude in region III, and there is no reflected wave.

Similarly, for an ingoing wave in region III we find
\begin{equation}
	{\rm Region~III \to Region~IV}: \qquad (r_{\rm c} -r)^{\frac{1}{2}
		- 2iq} \rightarrow -i e^{-2\pi q} \left|r- r_{\rm c}
		\right|^{\frac{1}{2} -2iq} \;, \label{io2}
\end{equation}
where we have taken $r_{\rm c} -r \rightarrow \left|
r-r_{\rm c}\right| e^{-\pi i}$.

Computing currents, we may calculate the transmission and reflection 
coefficients $T_{\rm c}$ and $R_{\rm c}$ of the corotation resonance:
\begin{equation}
	T_{\rm c} = e^{-4\pi q}, \quad R_{\rm c} = 0 \;. \label{tcrc}
\end{equation}
Thus, there is no wave reflection at the corotation singularity, and
there is a severe absorption of wave action in the transmitted
wave. Indeed, the absorption is exponentially strong when $q$ is large
(as for a thin disk).

From equation~(\ref{tcrc}), it might appear that the absorption would
disappear
for $n b^2 \le 1/4$ since then $q$ becomes zero or imaginary and so
$|T_{\rm c}| = 1$. However, this is not true in general, and
absorption disappears only when $n b^2 = 0$ (i.e., $q = \pm i/4$). For
example, consider a solution in region IV containing both ingoing and
outgoing waves
\begin{eqnarray}
	Q_r^{\rm (IV)} = A_1 (r-r_{\rm c})^{\frac{1}{2}+2iq} + A_2 (r-r_{\rm 
		c})^{\frac{1}{2}-2iq} \;,
\end{eqnarray}
where $A_1$ and $A_2$ are complex numbers. The corresponding solution in region
III obtained by analytic continuation is
\begin{eqnarray}
	Q_r^{\rm (III)} = i A_1 e^{-2\pi q}\left|r-r_{\rm c}\right|^{\frac{1}
		{2}+2iq} + i A_2 e^{2\pi q}\left|r-r_{\rm c}\right|^{\frac{1}
		{2}-2iq} \;.
\end{eqnarray}
Then we can calculate the net currents on the two sides of corotation. We 
find
\begin{equation}
        \frac{J_{\rm in}^{({\rm IV})} + J_{\rm out}^{({\rm IV})}}
        	{J_{\rm in}^{({\rm III})} + J_{\rm out}^{({\rm III})}} 
        	= -\frac{1}{\cos4\pi\overline{q} + \cot\psi\sin4\pi
		\overline{q}} \;,
        \label{nettc}
\end{equation}
where $\overline{q}\equiv iq$ and $2\psi\equiv\arg(A_1A_2^*/A_1^*A_2)$. The
ratio of the currents is equal to $1$ for any $\psi$ if and only if $q
= \pm i/4$, i.e.  $n b^2 = 0$ (in which case the corotation
singularity disappears or becomes an apparent singularity and we
return to the case studied by NGG).  Thus we conclude
that, in general, non-axisymmetric $g$-modes in disks are absorbed at
corotation.  In other words, the corotation singularity de-amplifies
waves and therefore cannot induce an instability.

Absorption at corotation is not unique to disk $g$-modes.  It occurs for 
other sorts of waves in shear flows and has been well studied in the
fluid-dynamics literature, e.g. \citet{drazin_reid}.  In particular,
a nonrotating, incompressible, stratified shear flow is absolutely linearly 
stable when the Richardson number
\[
R\equiv\frac{-g d\ln\rho/dx}{(dV_y/dx)^2}
\]
is greater than $1/4$ \citep{Howard61}.  Here $g$ is the acceleration
of gravity, $V_y$ the unperturbed velocity, and $x$ the vertical direction.  
\citet{Booker} studied $g$-modes in such flows as an initial value problem 
and showed that all but a fraction $\exp\left(-\pi\sqrt{R-1/4}\right)$ of
the wave current
is absorbed near the altitude $x_{\rm c}$ where $V(x_{\rm c})=\omega/k_y$, 
called the ``critical layer''.  They also showed that in the WKB 
approximation, wave packets propagate towards the critical layer but
fail to reach it in finite time. In the present problem, the role of the 
critical layer is played by corotation, and that of the Richardson number 
by $nb^2=n \kappa^2/\left(m h d\Omega/dr\right)^2$.

Combined with the results in \S\ref{sec5}, we see that, when $n>0$,
there are no amplifiers in the system, either at the two barriers or
at corotation, and so there are no non-axisymmetric growing modes
trapped in the disk. This result has been obtained here via a WKB
analysis.  We confirm the result in Appendix~\ref{appa} with numerical
calculations for the specific case of the shearing sheet model.  That
analysis is more general and goes beyond the WKB approximation.

\section{Summary and Discussion}
\label{sec7}

We have derived the wave equation for non-axisymmetric perturbations
in an isothermal hydrodynamic thin disk, and have identified a
conserved current (\S\ref{sec2}). The wave equation contains
two types of singularities: the corotation singularity and the
Lindblad singularity. The Lindblad
singularity is never a real singularity, regardless of the vertical
wave-number $n$; the solutions are always analytic and well-behaved in
the vicinity of this singularity, and the current is conserved
across it (\S\ref{sec3}).  For $n=0$ (horizontal motions independent of 
height), corotation is also not a real singularity, and the current is 
conserved.  But for waves having vertical nodes ($n>1$), corotation is 
a true singularity where the conservation of current breaks down.

Using a WKB approach (\S\ref{sec4}), we have shown that for $n=0$
there are two permitted regions in the disk
(Fig.~1), one near the inner boundary (region I), and the other at
large radii (region II).  In contrast, for the first and higher
overtones ($n>0$), there are four permitted regions in the disk (Fig.~2):
region I near the inner boundary, region II at large radii, and
regions III and IV in between, on either side of corotation.  We have
analyzed the WKB solutions in each permitted region, mapped the
ingoing and outgoing waves, and identified the signs of the current
density and group velocity of the various waves (summarized in
Table~\ref{tab1}).

For $n = 0$, the current is conserved
throughout the entire disk, from the inner boundary to infinity.
This, combined with the results given in Table~\ref{tab1}, allows us
to demonstrate that the disk behaves like a wave amplifier
(\S\ref{sec5}).  The argument is simple.  We consider an ingoing wave
in region I, which produces a reflected outgoing wave in region I and
a transmitted outgoing wave in region II.  Since the outgoing wave in
region II has a positive current,  by the conservation of
current, this means that the sum of the ingoing and outgoing
waves in region I should also have a positive current.  However, in
region I, the outgoing wave has a positive current and the ingoing
wave has a negative current.  Thus, the outgoing wave in region I must
have a larger amplitude than the ingoing wave.  In other words, the
interaction of the ingoing wave with the corotation region has caused
wave amplification in the reflected outgoing wave.

This result for $n=0$ waves was shown for the case of
the shearing sheet model by GN and NGG.  Indeed, 
these authors showed that the amplifier causes an instability if a 
reflecting boundary condition exists in one of the permitted regions.  
The reflective boundary causes the amplified outgoing wave to return to 
corotation 
and be amplified repeatedly. While this instability is a candidate to 
explain QPOs in accretion systems, it unfortunately requires nearly perfect 
reflection at the boundary, presumably
the inner edge of the disk, which is somewhat
problematic.  As the analytical results in the above-quoted papers
indicate, and also confirmed in the numerical calculations of
Appendix~A of the present paper (see the first two rows of Table~2),
the amplification at the corotation barrier is usually extremely weak
for a thin disk.  Thus, any energy loss either during transit of the
wave to the boundary and back, or during the reflection at the
boundary, would kill the instability.  The requirement of perfect
reflection at the boundary is particularly troublesome, since Blaes
(1987) has shown that any radial inflow of the gas at the boundary
(which is unavoidable in an accretion flow) would severely reduce the
reflectivity of the boundary.

What is needed is a
sufficiently strong amplifier, such that even less than perfect
reflection at the boundary is sufficient to give overall
amplification and instability.  Kato's (2001b, 2002) claim that
non-axisymmetric $g$-modes with $n>0$ are strongly amplified appeared
to be just the answer.  Unfortunately, Kato (2003) himself discovered
that $g$-modes do not grow, and our independent analysis described in
this paper confirms his result.  We have shown in \S\ref{sec4} that
for first and higher overtone modes ($n>0$), there are two barriers,
one near each Lindblad resonance, instead of the single barrier for
the $n=0$ case (compare Figs. 1 and 2).  Both barriers unfortunately
behave as wave de-amplifiers rather than amplifiers (\S\ref{sec5});
for a wave that is incident on either barrier from either side, both
the reflected wave and the transmitted wave are weaker than the
original wave (the total current is however conserved).
Furthermore, for these $n>0$ waves, the corotation singularity behaves
as a strong wave absorber (\S\ref{sec6}) so that current is
absorbed and is lost whenever a wave is incident on corotation.  Thus,
we conclude that $n>0$ modes are unable to grow in a thin disk.

If we define $g$-modes as those that have the bulk of their wave
action near corotation (Silbergleit et al. 2001; Kato 2002), then
these modes must of necessity have $n>0$, since only such modes have a
permitted region near corotation (Fig.~2).  Our analysis thus shows
that non-axisymmetric $g$-modes cannot be dynamically unstable.
$p$-modes, which have their wave action far away from corotation, are
possible both for $n=0$ and $n>0$.  Unstable $n>0$ $p$-modes are again
ruled out by our analysis.  This leaves only the $n=0$ modes, which we
discussed earlier, as candidates for QPOs.  We have already argued
that these modes are unlikely to grow in thin disks.  Perhaps with a
sufficiently thick disk, the wave amplifier could become strong (e.g.,
see the third row in Table~2) and might give an instability.  But this
is not very promising since it is likely that a thick disk will
also have more rapid gas inflow at the inner boundary and therefore
weaker reflection there.

Perhaps by adding more physics to the disk model one may yet find
a global hydrodynamic instability suited to explain QPO.  As already
noted, axisymmetric \emph{viscous} instabilities have been proposed
by \citet{ort00}.  Vertical stratification (i.e., perpendicular
to the mean flow) may allow non-axisymmetric instability in some
laboratory Couette flows \citep{Yavneh_etal01}.
In our opinion, neither of these
is a very plausible instability mechanism for disks,
but space does not permit an adequate discussion here.

The answer may lie with magnetohydrodynamic effects.  To
date, most detailed studies of magnetorotational instability have
emphasized the production of turbulence rather than high-$Q$
global oscillations, but we suspect that this is
the most promising direction for future work.

\acknowledgments 

LXL and RN thank the Institute for Advanced Study and the Department
of Astrophysical Sciences, Princeton, for hospitality while part of
this work was being done. LXL's research was supported by NASA through
Chandra Postdoctoral Fellowship grant number PF1-20018 awarded by the
Chandra X-ray Center, which is operated by the Smithsonian
Astrophysical Observatory for NASA under contract NAS8-39073.  RN's
research was supported in part by NSF grant AST-9820686 and NASA grant
NAG5-10780, and JG's by NASA grants NAG5-8385 and NAG5-1164.

\begin{appendix}
\section{The Shearing Sheet Model}
\label{appa}

In this Appendix we approximate a differentially rotating thin disk by 
means of the shearing sheet, which consists of a uniform shear flow with 
a Coriolis force (NGG). We use a Cartesian coordinate system $({x},
{y},{z})$, where ${x}$ and ${y}$ are related to the polar coordinates 
$r$ and $\phi$ of the original disk by
\begin{equation}
        {x} = r - r_{\rm c}\;, \quad {y} = r_{\rm
                c} (\phi - \Omega_{\rm c} t) \;. \label{xizeta}
        \label{xy}
\end{equation}
Here, $r_{ \rm c}$ is a reference radius and $\Omega_{\rm c} =
\Omega(r_{\rm c})$ is the disk angular velocity at $r_{\rm c}$. The
velocity of the unperturbed flow is ${\bf v} = 2 A {x} {\bf j}$ where
${\bf j}$ is a unit vector in the ${y}$-direction.  The frequency
$A$ and the related frequency $B$ are the Oort constants of the disk
at $r_{\rm c}$:
\begin{equation}
        A \equiv \left.\frac{r}{2} \frac{d\Omega}{d r}
                \right\vert_{r= r_{\rm c}} \;, \hspace{1cm}
        B \equiv \left.\frac{1}{2 r} \frac{d}{d r}\left(r^2
                \Omega\right)\right\vert_{r= r_{\rm c}} \;.
        \label{cacb}
\end{equation}
The epicyclic frequency of the flow in the ${x}{y}$-plane is
\begin{equation}
        \kappa \equiv \sqrt{4 B \Omega_{\rm c}} = \sqrt{4B (B-A)} \;.
        \label{epi}
\end{equation}
We treat the frequencies $A$, $B$, $\kappa$ as constants. 

When $\Omega \propto r^{-q}$ where $q$ is a constant, we have $2A = -q
\Omega_{\rm c}$, $2B = (2-q) \Omega_{\rm c}$, and $\kappa = \sqrt{2(2-q)}\,
\Omega_{\rm c}$. It is useful to remark that $q = 2$ corresponds to a disk
with constant angular momentum, $q = 3/2$ to a thin Keplerian disk, and
$q = 1$ to a disk with constant circular velocity. For $0<q \le 2$, 
we have $A < 0$ and $B \ge 0$.

We assume that the flow is isothermal with an equation of state
$p = \rho c_{\rm s}^2$, where $p$ is the gas pressure, $\rho$ is
the mass density, and $c_{\rm s} = {\rm constant}$ is the sound speed.
We take the flow to be homogeneous and to extend to $\pm\infty$ along
${x}$ and ${y}$.  In the vertical direction we make the usual
harmonic approximation for the potential, with vertical frequency
$\Omega_\perp$.  Then, vertical hydrostatic equilibrium implies
$p\propto\rho\propto \exp(-{z}^2/2 h^2)$, where the vertical scale
height $h$ is equal to $c_{\rm s}/\Omega_\perp$.  In a thin Keplerian
disk, $\Omega_\perp=\Omega_c=\kappa$, but we allow these three
frequencies to be different.

Without loss of generality, we assume linear perturbations proportional 
to $\exp[i(- \omega t +k_{y} {y})]$, where $k_{y}=m/r_{\rm c}$ is 
the azimuthal wave-vector, and the frequency $\omega$ may in general be 
complex. Then, the first-order perturbation equations are
\begin{eqnarray}
	\frac{\partial u}{\partial {x}} + i k_{y} v + \frac{1}{\rho_0}
		\frac{\partial}{\partial {z}}(\rho_0 w) &=& \frac{i
		\sigma}{c_{\rm s}^2} Q \;,\label{e1} \\
	i\sigma u + 2\Omega_{\rm c} v &=& \frac{\partial Q}{
		\partial {x}} \;,\label{e2} \\
	2B u -i\sigma v &=& -ik_{y} Q \;,\label{e3} \\
	i\sigma w &=& \frac{\partial Q}{\partial {z}} \;,\label{e4}
\end{eqnarray} 
where $u$, $v$, and $w$ are the perturbations to the velocities in the 
${x}$, ${y}$, and ${z}$ directions respectively, $\rho_0$ is the unperturbed
mass density, $Q \equiv \delta p/\rho_0$ is the perturbed enthalpy, and
\begin{equation}
        \sigma \equiv \omega - m \Omega (r) =- 2 k_{y} A{x} 
        \label{sig}
\end{equation}
is the pattern frequency of the mode, where we have chosen $r_{\rm c}$
to be the corotation radius where the fluid moves with the same speed as 
the pattern speed of the mode
\begin{equation}
        m \Omega(r_{\rm c}) = \omega \;.
        \label{rc}
\end{equation}
Equations~(\ref{e1}--\ref{e4}) correspond to equations~(\ref{cont}) and
(\ref{rmom}--\ref{zmom}) for the case of a thin disk, respectively.

From equations~(\ref{e1}--\ref{e4}) a second-order partial differential 
equation in ${x}$ and ${z}$ may be derived for any of the perturbed flow 
quantities.  Following Kato (2002), we consider the equation for $Q$. The
isothermal equation of state allows a separation of variables, $Q= Q_x({x}) 
H_n({z}/\sqrt{2}h)$, where $H_n(\eta)$ is a Hermite polynomial of degree 
$n\ge0$ (see Okazaki et al. 1987 and Kato 2002 for details). The function 
$Q_x({x})$ then satisfies the following differential equation:
\begin{equation}
        \frac{d}{d {x}}\left(\frac{1}{D} \frac{d Q_x}{d {x}}\right) +
                \left(\frac{n \Omega_\perp^2 - \sigma^2}{c_{\rm s}^2
                \sigma^2} - \frac{k_{y}^2}{D} - 
		\frac{2 k_{y} \Omega_{\rm c}}{\sigma}
                \frac{d}{d {x}} \frac{1}{D}\right)
                Q_x = 0 \;.
        \label{df0}
\end{equation}
where $D \equiv \kappa^2 - \sigma^2$. 
Equation~(\ref{df0}) is similar to equation~(\ref{deq1}) which we derived for
the thin disk. In the same way, we can define a conserved current 
by
\begin{eqnarray}
        J \equiv \frac{i}{2 D} \left(Q_x^*\, \frac{dQ_x}{d{x}} -
                Q_x\, \frac{dQ_x^*}{d{x}}\right) \;. 
        \label{curr}
\end{eqnarray}

As in a thin disk, there are two types of singularities in 
equation~(\ref{df0}): the Lindblad singularity at $D= 0$, and the corotation
singularity at $\sigma = 0$. The Lindblad singularity is always an apparent 
singularity. The corotation singularity is a true singularity when $n >0$ 
but apparent
when $n = 0$. In the WKB limit, there are four permitted regions in a 
shearing sheet flow when $n>0$: region I with $\sigma < -\max(\sqrt{n}
\Omega_\perp,\kappa)$, region II with $\sigma > \max(\sqrt{n}\Omega_\perp,
\kappa)$, region III with $-\min(\sqrt{n}\Omega_\perp,\kappa) <\sigma <0$, 
and region IV with $0<\sigma <\min(\sqrt{n}\Omega_\perp,\kappa)$. Region I 
(II) and region III (IV) are separated from each other by a potential barrier, 
region III and region IV are separated from each other by the corotation 
singularity. When $n = 0$, there are only two permitted regions: I and II. 

The results in the main text can be applied to the shearing sheet model. In
particular, when $n = 0$, the potential barrier between regions I and II 
behaves as an amplifier, a wave incident on it is reflected with larger
amplitude (NGG). When $n>0$, the two potential barriers (one between
regions I and III, the other between regions II and IV) behave as a 
de-amplifier, a wave incident on it is reflected with smaller amplitude. In
addition, when $n >0$, the corotation singularity absorbs energy from a wave. 

Here we show some numerical results for the shearing sheet model to confirm 
these results. Since these numerical solutions do not require that the WKB 
approximation be valid in regions III and IV, they complement the results 
obtained in the main text.

For numerical work, it is convenient to replace equation~(\ref{df0}) with
an equivalent set of two first-order differential equations. Assuming as 
before that all variables depend upon $z$ as Hermite polynomials 
$H_n(z/\sqrt{2}h)$ or its derivative,
we eliminate $w$ and $v$ from 
equations~(\ref{e1})--(\ref{e4}) to obtain, in matrix form,
\begin{equation}
\frac{d}{dx}\left[\begin{array}{c}u_x\\[1ex] Q_x\end{array}\right] =
\frac{1}{\sigma}\left[
\begin{array}{c@{\qquad}c}
-2 k_{y} B & i(\sigma^2 -n\Omega_\perp^2 -k_{y}^2 c_{\rm s}^2)/c_{\rm s}^2
\\[1ex]
i(\sigma^2-\kappa^2) & 2 k_{y}\Omega_{\rm c}
\end{array}\right]\left[\begin{array}{c}u_x\\[1ex] Q_x\end{array}\right].
\label{cemat}
\end{equation}
In this first-order form of the linearized equations, it is clear that
the only possible singularity is $\sigma=0$.  This explains why the
Lindblad resonances $\sigma=\pm\kappa$ are not true singularities of
the equivalent second-order equations~(\ref{df0}) and (\ref{deq1}).

We look for solutions to equations~(\ref{cemat}) that contain
only outgoing waves in region II. To do so, we start from an outgoing wave at
$x_0 >0$ in the WKB region II, which has a positive wave-vector (thus a 
positive group velocity), then integrate the equations along the real axis of 
$+x$ toward the corotation. Near but before crossing the corotation, we deform 
the integration path into the complex plane of $x$ to pass the corotation 
singularity from 
above, then come back to the real axis of $-x$. Then, we integrate the 
equations along the real axis of $-x$ into the WKB region I, until $x = -x_0$ 
is reached. At $x = -x_0$, we decompose the solution into an ingoing component 
(having a negative wave-vector, i.e., a positive group velocity) and an 
outgoing component (having a positive wave-vector, i.e., a negative group 
velocity). Then, we calculate the gain $G$ (i.e., the reflection coefficient), 
and the transmission coefficient $T$ by
\begin{eqnarray}
	G &=& \left|\frac{u_{x,{\rm out}}(x=-x_0)}{u_{x,{\rm 
		in}}(x = -x_0)}\right|^2 = \left|\frac{Q_{x,{\rm 
		out}}(x=-x_0)}{Q_{x,{\rm in}}(x = -x_0)}
		\right|^2 \;,
	\label{gain} \\
	T &=& \left|\frac{u_{x,{\rm out}}(x=x_0)}{u_{x,{\rm 
		in}}(x = -x_0)}\right|^2 = \left|\frac{Q_{x,{\rm 
		out}}(x=x_0)}{Q_{x,{\rm in}}(x = -x_0)}
		\right|^2 \;,
	\label{trans} 
\end{eqnarray}
where ``out'' denotes ``outgoing'', ``in'' denotes ``ingoing''.

Numerical results corresponding to a few choices of parameters are shown
in Table~\ref{tab2}. They are classified into two classes: one with $n = 0$; 
the other with $n = 1$. Each class contains both a Keplerian disk ($\Omega_\perp 
= \kappa = \Omega_{\rm c}$) and non-Keplerian disks ($\Omega_\perp$ and 
$\kappa$ different from $\Omega_{\rm c}$). The results show that, for the case 
of $n=0$, where the corotation is not an intrinsic singularity, the incident 
wave is amplified: $G > 1$; indeed $G = 1 + T$ since the current is conserved. 
One can check that the numerical results agree with the analytical results 
of NGG, where
\begin{eqnarray}
	G = 1 + T = 1 + \exp(-2\pi C) \;, \hspace{1cm}
	C = \frac{c_{\rm s}^2 k_{y}^2 +\kappa^2}{4 c_{\rm s}|A 
		k_{y}|} \;.
\end{eqnarray}
However, for the case of $n = 1$, where the corotation is an intrinsic 
singularity, there is strong absorption at the corotation as indicated by the 
fact that $T \ll 1- G$, and
the incident wave is always de-amplified since $G < 1$. 

The initial and terminal points of our numerical integrations are always
chosen well within the WKB domain of regions II and I, respectively
(i.e., $|k_x x|\gg 1$, where $k_x$ is the wave-vector in the $x$-direction), 
so that the ingoing and outgoing waves are clearly distinguishable.
It can be shown that for $n\ge1$, the 
criterion for the validity of the WKB approximation in regions III and IV
near corotation reduces to $b \gg 1$, where
\begin{eqnarray}
	b\equiv \frac{\kappa}{-2A k_{y} h} \;. \nonumber
\end{eqnarray}
The solutions listed in Table~\ref{tab2} do not satisfy $b\gg1$.
Indeed, for the last solution, where $\Omega_\perp 
= 0.8 \Omega_{\rm c}$, $\kappa = 0.2 \Omega_{\rm c}$, and $k_{y} h = 0.3$, 
we have $b = 0.336$ and $q = (1/2)\sqrt{nb^2 -1/4} = 0.185 i$. Even for this 
case with an imaginary $q$, we see that $T < 1-G$, which
must be attributed to absorption at corotation.
These numerical results support and complement our 
analytical results based on the WKB approximation.  

\end{appendix}


\clearpage
\begin{deluxetable}{lcccc}
\tablewidth{0pt}
\tablecaption{Relations among the directions of the current, 
the wave-vector, and the group velocity in the four permitted regions 
\label{tab1}}
\tablehead{
\colhead{Region} & \colhead{$D\equiv \kappa^2-\sigma^2$} &
\colhead{~~~~$k_r J$\tablenotemark{a}~~~~} & \colhead{~~~~$k_r v_{{\rm g}
r}$\tablenotemark{b}~~~~}
& \colhead{~~~~$v_{{\rm g}r} J$\tablenotemark{c}~~~~}
}
\startdata
I   & $-$   & $+$   & $-$   & $-$  \\
II  & $-$   & $+$   & $+$   & $+$  \\
III & $+$   & $-$   & $+$   & $-$  \\
IV  & $+$   & $-$   & $-$   & $+$  \\
\enddata

\tablenotetext{a}{The sign of the product $k_r J$ determines the relative 
directions between the current $J$ and the wave-vector $k_r$.}
\tablenotetext{b}{The sign of the product $k_r v_{{\rm g} r}$ determines 
the relative directions between the wave-vector $k_r$ and the group velocity
$v_{{\rm g} r}$.}
\tablenotetext{c}{The sign of the product $v_{{\rm g}r} J$ determines the 
relative directions between the current $J$ and the group velocity
$v_{{\rm g} r}$.}

\tablecomments{When $n = 0$ regions III and IV disappear 
(see \S\ref{sec4}).}

\end{deluxetable}

\clearpage
\begin{deluxetable}{ccccccc}
\tablewidth{0pt}
\tablecaption{Numerical solutions for gain and transmission for the
shearing sheet model
\label{tab2}}
\tablehead{
\colhead{$n$} & \colhead{$\Omega_\perp/\Omega_{\rm c}$} &
\colhead{$\kappa/\Omega_{\rm c}$} & \colhead{~$k_{y} h$}~~
& \colhead{~$G$\tablenotemark{a}}~ & \colhead{~~$T$\tablenotemark{b}}
& \colhead{Amplification\tablenotemark{c}}
}
\startdata
$0$   & $1.0$   & $1.0$   & ~$0.3$~~   & ~$1.000496$~  & ~~$4.96$E$-4$ & $\surd$ \\
$0$   & $1.0$   & $1.1$   & ~$0.4$~~   & ~$1.000447$~  & ~~$4.47$E$-4$ & $\surd$ \\
$0$   & $0.8$   & $0.2$   & ~$0.3$~~   & ~$1.524428$~  & ~~$5.24$E$-1$ & $\surd$ \\
$1$   & $1.0$   & $1.0$   & ~$0.3$~~   & ~$3.89$E$-1$~ & ~~$4.60$E$-7$ & $\times$ \\
$1$   & $1.0$   & $1.1$   & ~$0.4$~~   & ~$4.94$E$-1$~ & ~~$1.60$E$-6$ & $\times$ \\
$1$   & $0.8$   & $0.2$   & ~$0.3$~~   & ~$9.38$E$-1$~ & ~~$7.63$E$-3$ & $\times$ \\
\enddata

\tablenotetext{a}{The gain, i.e., the reflection coefficient, defined by 
eq.~(\ref{gain}).}
\tablenotetext{b}{The transmission coefficient defined by eq.~(\ref{trans}).}
\tablenotetext{c}{Amplification ($G>1$) is marked with ``$\surd$'';
De-amplification ($G<1$) is marked with ``$\times$''.}


\end{deluxetable}

\clearpage

\clearpage

\end{document}